\documentclass[11pt]{article}
\long\def\symbolfootnotemark[#1]{\begingroup%
\def\thefootnote{\fnsymbol{footnote}}\footnotemark[#1]\endgroup}
\long\def\symbolfootnotetext[#1]#2{\begingroup%
\def\thefootnote{\fnsymbol{footnote}}\footnotetext[#1]{#2}\endgroup}
\topmargin = - 1.5 cm
\usepackage{amsmath}
\usepackage{amssymb}

\newtheorem{Theorem}{Theorem}
\newtheorem{Dogma}{Dogma}
\newtheorem{Proposition}{Proposition}

\newcommand{\Hs}{Hilbert space} 
\newcommand{\HN}{H^{\otimes N}}
\newcommand{\vna}{von Neumann algebra}
 
\newcommand{\al}{\alpha} 
 
 \newcommand{\Dl}{\Delta}

\newcommand{\lm}{\lambda} 
 
 \newcommand{\ta}{\tau} 
 \newcommand{\phv}{\varphi}
 
\newcommand{\ot}{\otimes} 
\newcommand{\half}{\mbox{\footnotesize $\frac{1}{2}$}}
\newcommand{\third}{\mbox{\footnotesize $\frac{1}{3}$}}

\newcommand{\raw}{\rightarrow}
\newcommand{\x}{\times} 
\newcommand{\ovl}{\overline}
\newcommand{\CA}{\mathcal{A}}

\newcommand{\R}{\mathbb{R}}
\newcommand{\C}{\mathbb{C}}
\newcommand{\Z}{\mathbb{Z}}
\newcommand{\cin}{C^{\infty}} \newcommand{\cci}{C^{\infty}_c}
\newcommand{\er}{\eqref}

 \newcommand{\rep}{representation}
\newcommand{\irrep}{irreducible representation}
\newcommand{\inv}{^{-1}} 
 \newcommand{\sg}{\sigma}
 \newcommand{\ch}{\chi} \newcommand{\ps}{\psi} 
\newcommand{\til}{\tilde}
\begin{document}
\setlength{\baselineskip}{1\baselineskip}
\thispagestyle{empty}
\title{Quantization and superselection sectors {\sc iii}: \\ Multiply connected spaces and indistinguishable particles}
\author{
 N.P. (Klaas) Landsman\footnote{Radboud University Nijmegen, Institute for Mathematics, Astrophysics, and Particle Physics,  Heyendaalseweg 135, 6562 {\sc aj} Nijmegen, The Netherlands.    Email: \texttt{landsman@math.ru.nl}. Twitter: \texttt{@Nulvolgers}.}}
 \maketitle
\begin{center}{\it
Dedicated to John Earman}
\end{center}
\bigskip

 \begin{abstract}
 We reconsider the (non-relativistic) quantum theory of indistinguishable particles on the basis of Rieffel's notion of C*-algebraic (`strict')  deformation quantization. Using this
  formalism, we  relate the operator  approach of  Messiah and Greenberg (1964) to the configuration space approach pioneered by Souriau (1967), Laidlaw and DeWitt-Morette (1971), Leinaas and Myrheim (1977), and others. In  the former, 
 the algebra of observables $\mathcal{M}_N$ of $N$ indistinguishable particles is postulated to be the $S_N$-invariant part of the corresponding  algebra describing $N$ distinguishable (but identical) particles (where $S_N$ is the permutation group on $N$ objects). According to the algebraic theory of superselection sectors,
irreducible \rep s of $\mathcal{M}_N$  then correspond to particle states of given permutation symmetry type, which in turn are labeled by arbitrary
 irreducible \rep s of $S_N$. Hence this approach yields   bosons, fermions, and paraparticles. In the latter approach, the classical configuration space  of $N$ indistinguishable (and impenetrable) particles in $Q=\R^d$ is postulated to be $Q_N=(Q^N-\Delta_N)/S_N$, where $\Delta_N$ is the appropriate $N$-fold generalization of the diagonal in $Q^N$.  Various  arguments involving covering spaces and line bundles then lead to the conclusion that physical wave-functions transform under some one-dimensional unitary \rep\ of the first homotopy group $\pi_1(Q_N)$. For $d>2$ this group equals $S_N$, leaving room for bosons and fermions only. This contradicts the operator approach as far as the admissibility of parastatistics is concerned.
 
  To resolve this, we first prove that in $d>2$  the topologically nontrivial configuration spaces of the second approach are quantized by the algebras of observables of the first. Second,  we show that the irreducible \rep s of the latter may be realized by vector bundle
constructions, which include the line bundles of  the second approach:
 \rep s on higher-dimensional bundles (which define parastatistics) cannot be excluded.
However, we show that the corresponding particle states may always be realized in terms of 
bosons and/or fermions with an unobserved internal degree of freedom.

 Although based on
 non-relativistic quantum mechanics, this conclusion is analogous to the rigorous results of the Doplicher--Haag--Roberts analysis in algebraic quantum field theory, as well as to  the heuristic arguments which led Gell-Mann and others to {\sc qcd}.
 \end{abstract}
 \newpage
 \section{Introduction}\label{S1}
 This is the third and (probably) last paper in our series on quantization and superselection rules, the first installments of which were published in 1990 \cite{landsman1990b,landsman1990c}. The present paper, which can be read independently of its predecessors, once more combines  C*-algebraic quantization theory with the algebraic theory of superselection sectors \cite{Haag}  (removed from its original context of quantum field theory),
now applying this combination to the theory of permutation invariance and indistinguishable particles. Doing so appears to be a new application of Rieffel's strict deformation quantization \cite{Rieffel89,Rieffel90,Rieffel94} (see also \cite{book,handbook}), and as a bonus we will be able to clarify the relationship between the two main theoretical treatments of permutation invariance so far, namely the 
  operator approach of  Messiah and Greenberg \cite{MG1964} and the configuration space approach due to Souriau \cite{Souriau1, Souriau2} and Laidlaw \& DeWitt-Morette \cite{LDW}. Indeed, these treatments turn out to be equivalent at least in dimension $d>2$.\footnote{In lower dimension this equivalence falters because of the difference between the  permutation group $S_N$
 used in the operator approach and the braid group $B_N$ appearing in the configuration space approach; cf.\ \S\ref{S9}.}
  
The issues of permutation invariance and indistinguishability arose explicitly in quantum statistical mechanics, but with hindsight the problem was already implicit in the Gibbs paradox of classical statistical mechanics;\footnote{Our take on this is the same as Heisenberg's and Dirac's (against Einstein):  at least in this case, it is not the theory which decides what is observable, but \emph{vice versa}. So if, in classical statistical mechanics, possessing the capability of observation of  a Laplacian Demon one is able to keep track of individual particles, one should use the phase space of $N$ distinguishable particles and hence Boltzmann's counting procedure. This leads to the Gibbs paradox \cite{SS} only if one then inconsistently assumes that alas one has suddenly lost track of the individual particles. If, more realistically, one's observables are permutation-invariant, then, as  argued by Gibbs himself \cite{Gibbs}, one should use the  state space of $N$ indistinguishable particles, treated also in the main body of the present paper. Hence it is the choice of the theoretical description that determines the counting procedure and hence the entropy (which therefore comes out as an intersubjective quantity). A fundamental difference between classical and quantum physics is that in the latter one does not have this choice: identical particles are necessarily permutation invariant. But this symmetry is broken by measurement, which individuates particles and recovers the ambiguous situation of classical physics just discussed.
}  see \cite{FrenchKrause}, and more briefly also  \cite{Bach,SS}. In sum, after a period of confusion but growing insight, involving some of the greatest physicists  such as Planck, Einstein,  Ehrenfest, Heisenberg, and Fermi, the modern point of view on quantum statistics
 was introduced by Dirac \cite[\S3]{Dirac1926}. Using modern notation, and abstracting from his specific example (which involved electronic wave-functions), his argument is as follows.\footnote{The permutation group $S_N$ ($\equiv$ symmetric group) was not explicitly used by Dirac; it was  introduced in quantum physics by Wigner in the following year \cite{Wigner1927}, following a suggestion of his friend von Neumann \cite[\S3.4.(c)]{MR6}. Note that Dirac's  mathematical formulation of permutation invariance and indistinguishability, like all later ones,  is  predicated on the possibility of initially making sense of identical yet distinguishable systems.  These can be subsequently be permuted as a nontrivial operation, invariance under which will eventually define indistinguishability.  In other words, in order to define (at least mathematically) what it means for identical systems  to be indistinguishable,
we must first  be able to describe them as distinguishable! } 

Let $H$ be the \Hs\ of a single quantum system, called a `particle' in what follows. The two-fold tensor product $H^{\ot 2}\equiv H\otimes H$ then describes two distinguishable copies of this particle. The permutation group $S_2$ on two objects,  with nontrivial element $(12)$, acts on the state space $H^{\ot 2}$ in the natural way, i.e., by linear extension of $U(12) \ps_1\ot\ps_2=\ps_2\ot\ps_1$. Praising Heisenbergs emphasis on defining everything in terms of observable quantities only, 
but unwittingly  echoing Leibniz's \emph{Principle of the Identity of Indiscernibles} ({\sc pii}),\footnote{This principle states that  two different objects cannot have exactly the same properties. In other words,  
 two objects that have exactly the same properties must be identical.} Dirac then declares the two particles to be indistinguishable if the observable quantities are represented by operators $A$ that are invariant under particle exchange, which we would now formulate as invariance under the induced action of $(12)$ on the operators, i.e., if $U(12)AU(12)^*=A$. 
 
 \noindent By unitarity, this is to say that $A$
 commutes with  $U(12)$. Dirac notes that such operators 
map symmetrized vectors (i.e. those $\psi\in H\otimes H$ for which $U(12) \ps=\ps$) into symmetrized vectors, and likewise map 
anti-symmetrized vectors (i.e. those $\psi\in H\otimes H$ for which $U(12) \ps=-\ps$) into anti-symmetrized vectors, and these are the only possibilities; we would now say that under the  action of the
$S_2$-invariant (bounded) operators one has
\begin{equation}
H^{\ot 2}\cong H^{\ot 2}_S\oplus H^{\ot 2}_A,\label{HSHA}
\end{equation}
 where 
\begin{eqnarray}
H_S^{\ot 2}&=&\{\ps\in H^{\ot 2}\mid U(12) \ps=\ps\} ; \label{defHS}\\
H_A^{\ot 2}&=&\{\ps\in H^{\ot 2}\mid U(12) \ps=-\ps\}. \label{defHA}
\end{eqnarray}
Arguing that in order to avoid double counting (in that $\ps$ and $U(12) \ps$ should not both occur as independent states) one has to pick one of these two possibilities, Dirac concludes that state vectors of a system of two indistinguishable particles must be either symmetric or anti-symmetric. Without further ado, he then generalizes this to systems of $N$ identical particles: if $(ij)$ is the element of the permutation group $S_N$ on $N$ objects that permutes $i$ and $j$ ($i,j =1, \ldots, N$), then according to Dirac, $\ps\in H^{\ot N}$ should satisfy either $U(ij) \ps=\ps$, in which case $\ps\in H^{\ot 2}_S$, or $U(ij) \ps=-\ps$, in which case $\ps\in H^{\ot 2}_A$. Here $U$ is the natural unitary \rep\ of $S_N$ on $H^{\ot N}$, given, on $\pi\in S_N$, by linear extension of
\begin{equation}
U(\pi) \ps_1\ot\cdots\ot \ps_N=\ps_{\pi(1)}\ot\cdots\ot \ps_{\pi(N)}. \label{defU}
\end{equation}
 Equivalently, 
$\ps\in H_S^{\ot 2}$ if it is invariant under all permutations, and $\ps\in H_A^{\ot 2}$ if it is invariant under all even permutations and picks up a minus sign under all odd permutations. This (non) argument leaves room for bosons and fermions alone, although the orthogonal complement of $H_S^{\ot 2}\oplus H_A^{\ot 2}$ in $H_N$ (describing particles with `parastatistics') is non-zero as soon as $N>2$. 

Since Nature has proved Dirac's feeble arguments to be right so far, much of the subsequent research on indistinguishable particles has had the goal of explaining away  the possibility of parastatistics, at least in dimension $d>2$.\footnote{In a  philosophical direction, some  interesting research has been concerned with logical questions of identity and with the possibility of permutation invariance in classical and quantum physics \emph{per se}, cf.\ \cite{CB,FrenchKrause,SS}.}  Although our results have some implications for that discussion,
our main goal is to clarify the relationship between the two main approaches to permutation invariance taken in the literature so far (cf.\ \cite{Earman}). Distinguished by the  different natural actions of $S_N$ they depart from, these are based on:
\begin{itemize}
\item \textbf{Quantum observables.}
$S_N$ acts on the (von Neumann)  algebra $B(\HN)$ of bounded operators on $\HN$ by conjugation of the unitary \rep\ $U(S_N)$ on $\HN$.
\item \textbf{Classical states.} $S_N$ acts on $M^N$, the $N$-fold cartesian product of the classical one-particle phase space $M$, by permutation. If $M=T^*Q$ for some configuration space $Q$, we might as well start from the natural action of $S_N$ on $Q^N$ (pulled back to $M^N$), and this is indeed what we shall do, often further simplifying to $Q=\R^d$.
\end{itemize}
This begs the question why we do not consider the action $U$ of $S_N$ on $\HN$ itself as a  starting point; the answer is that taken by itself, this is  a non-starter. Omitting references in grace,\footnote{See \cite{Earman} for an extensive list, including even such giants as Landau and Lifshitz, \emph{Quantum Mechanics}.} authors who try to derive the bose/fermi alternative this way typically reason as follows:
\begin{quote}
`Since, in the case of indistinguishable particles, $\ps\in\HN$ and $U(\pi)\ps$ must represent  the same state for any $\pi\in S_N$, and since two unit vectors represent the same state iff they differ by a phase vector, it must be  that $U(\pi)\ps=c(\pi)\ps$, for some $c(\pi)\in\C$ satisfying $|c(\pi)|=1$ (by unitarity). The group property $U(\pi\pi')=U(\pi)U(\pi')$ then implies that $c(\pi)=1$ for  even permutations and $c(\pi)=\pm 1$ for  odd permutations. The choice $+1$ in the latter leads to bosons, whereas $-1$ leads to fermions, so these are the only possibilities.'
\end{quote} 
Despite its popular appeal, this argument is either incomplete or at best circular \cite{Earman}:
\begin{itemize}
\item The phase vector $c(\pi)$ might depend on $\ps$;
\item More importantly, the claim that two unit vectors represent the same state iff they differ by a phase vector, presumes  that the particles are distinguishable!
\end{itemize}
Indeed, the only argument that two unit vectors $\ps$ and $\ps'$ are equivalent iff $\ps'=z\ps$ with $|z|=1$, is that it guarantees that expectation values coincide, i.e., that $(\ps,A\ps)=(\ps',A\ps')$ for all (bounded) operators $A$.
But, following Heisenberg and Dirac,  the whole point of having indistinguishable particles is that an operator $A$ represents a physical observable iff it is invariant under   all permutations (acting by conjugation). Requiring $(\ps,A\ps)=(\ps',A\ps')$ only for such operators leaves far more possibilities, as we shall see in the next section.

The two remaining approaches  above have  developed independently. The former goes back (at least)  to Messiah and Greenberg  \cite{MG1964}, whereas the latter was independently  introduced  by Souriau \cite{Souriau1,Souriau2} and 
 by Laidlaw \& DeWitt-Morette \cite{LDW}. Often in the wider context of the quantization of multiply connected spaces, it was subsequently developed in various ways \cite{Dowker,Horvathy1,Horvathy2,HorvathyMorandiSudarshan,SudarshanImboImbo,Isham84,LM1,SchulmanBook}, all of which give essentially the same result, viz.\ eq.\ \er{equi1} in \S\ref{S3}.
 
Our aim is to relate  these approaches, towards which goal we proceed as follows. 
In \S\ref{S2} we review the operator approach (in modern form). Section \ref{S3} reviews the configuration space approach to indistinguishability, but here we feel the need of a slightly more critical approach, triggered by the  major discrepancy between the two approaches in question: the former admits parastatistics but the second (apparently) does not. However, this apparent conclusion turns out to be based on an unnecessary self-imposed limitation to scalar wave-functions  in the step of passing from configuration spaces or phase spaces to \Hs s (sometimes caused by essentially the same mistake as the one just pointed out above).

This clears the way for a unification of both methods on the basis of (strict) deformation quantization, whose outspoken goal is precisely to relate operator algebras to classical phase spaces in a systematic way. Our general framework is reviewed in \S\ref{S4}, and is subsequently applied to an illuminating example, namely the quantization of spin, in \S\ref{S5}. This example, which contains the main feature of our full problem, namely a multiple connected phase space, in an embryonic way, is then generalized from Lie groups to Lie groupoids (cf.\  \S\ref{Intermezzo}). This
paves the way for the strict deformation quantization of arbitrary multiply connected spaces in \S\ref{S6}.
As (another) warm-up we first apply our formalism to a particle moving on a circle (\S\ref{S7}), before coming to the  quantization of indistinguishable particles in \S\ref{S8}.
We close with a discussion in \S\ref{S9}, addressing among others  the role of quantum field theory.
\subsection*{Acknowledgement}
This paper owes its existence to Hans Halvorson, who drew the attention of the author to the possibility of applying his earlier work on quantization to permutation invariance and indistinguishability. The author's views on these issues were considerably  influenced by John Earman, whose unpublished paper \cite{Earman} set the stage for the present one. 
\section{Permutation-invariant quantum observables}\label{S2}
Here one implements permutation invariance (in the Heisenberg--Dirac way) by postulating that the physical observables of the $N$-particle system under consideration be the $S_N$-invariant operators: with $U$  given by \er{defU}, the algebra of observables is taken to be 
\begin{equation}
 \mathcal{M}_N=B(\HN)^{S_N}\equiv \{A\in B(\HN)\mid [A,U(\pi)]=0\,\forall \pi\in S_N\}. \label{defMN}
\end{equation}
Since it has the same reduction on $\HN$, for our purposes one may alternatively take 
\begin{equation}
 \mathcal{A}_N=K(\HN)^{S_N}\equiv \{A\in K(\HN)\mid [A,U(\pi)]=0\,\forall \pi\in S_N\},
\end{equation}
where $K(\cdot)$ are the compact operators, so that \er{defMN} is just the bicommutant $\mathcal{M}_N=\mathcal{A}_N''$. Yet another perspective on $\mathcal{M}_N$ is that it is the \vna\ $(U(H)^{\ot N})''$ generated by the $N$-fold tensor product $U(H)^{\ot N}$ of the defining \rep\ of the group of unitary operators $U(H)$ on $H$.\footnote{If $H\cong \C^m$ one simply has the usual unitary matrix group $U(H)=U(m)$; for infinite-dimensional $H$ one needs the right topology on $H$ to define $U(H)$, as discussed e.g.\ in \cite{Landsman1999}.}  
As long as $\dim(H)>1$ and $N>1$, the algebras $\mathcal{M}_N$ and $\mathcal{A}_N$ act reducibly on $\HN$. 
The reduction of $\HN$ under $\mathcal{M}_N$ (and hence of $\mathcal{A}_N$ and of $U(H)^{\ot N}$)
 is easily carried out by  Schur duality \cite{GoodmanWallach}. A \emph{partition}
$\lm$ of $N$ is a way of writing 
\begin{equation}
N=n_1+\cdots + n_k, \:\: n_1\geq \cdots \geq n_k>0, \:\: k=1,\ldots, N, \label{partition}
\end{equation}
with corresponding \emph{frame} $F_{\lm}$, which is simply a picture $N$ boxes with $n_i$ boxes in the $i$'th row, $i=1,\ldots, k$. For each frame $F_{\lm}$, one has $N!$ possible \emph{Young tableaux} $T$, each of which is a particular way of  writing all of the numbers 1 to $N$ into the boxes of $F_{\lm}$.
A Young tableau is \emph{standard} if  the entries in each row increase from left to right and the entries in each column increase from top to bottom.  The set of all (standard) Young tableaux on $F_{\lm}$ is called 
$\mathcal{T}_{\lm}$ ($\mathcal{T}^S_{\lm}$).  Clearly, $S_N$ acts on $\mathcal{T}_{\lm}$ in the obvious way by permutation.
To each $T\in \mathcal{T}_{\lm}$ we associate the  subgroup $\mathrm{Row}(T)\subset S_N$  of all $\pi\in S_N$ that preserve each row (i.e., each row of $T$ is permuted within itself) and likewise
 $\mathrm{Col}(T)\subset S_N$ consists of all $\pi\in S_N$ that preserve each column. 
The set $\mathrm{Par}(N)$ of all partitions $\lm$ of $N$ parametrizes the conjugacy classes of $S_N$ and hence also the (unitary) dual of $S_N$; in other words, up to (unitary) equivalence each (unitary) \irrep s $U_{\lm}$ of $S_N$ bijectively corresponds to some partition $\lm$ of $N$;
the dimension of any vector space $V_{\lm}$ carrying $U_{\lm}$ is $N_{\lm}=|\mathcal{T}^S_{\lm}|$, the number of different standard Young tableaux on $F_{\lm}$. 
Returning to \er{defU}, to each $\lm\in\mathrm{Par}(N)$ and each Young tableau  $T\in \mathcal{T}_{\lm}$ we associate an operator $P_T$ on $\HN$ by the formula
\begin{equation}
P_T=\frac{N_{\lm}}{N!}\sum_{\pi\in\mathrm{Col}(T)}\mathrm{sgn}(\pi)U(\pi)\sum_{\pi'\in\mathrm{Row}(T)}U(\pi'),
\end{equation}
which happens to be a projection. Its image $P_T\HN\subset\HN$ is denoted by $\HN_T$, and the restriction of $\mathcal{M}_N$ to $\HN_T$ is called $\mathcal{M}_N(T)$.
 One may now write the decomposition of $\HN$ under the reducible defining action of $\mathcal{M}_N$  
   in two slightly different ways, each with its own merits. Assuming  $\dim(H)\geq N$,\footnote{If $\dim(H)<N$, then only partitions \er{partition} with $k\leq \dim(H)$  occur in \er{decN1} and \er{decN2}.} $\HN$ decomposes into irreducibles under $\mathcal{M}_N$ (etc.) as 
\begin{eqnarray}
\HN&=& \bigoplus_{T\in\mathcal{T}^S_{\lm},\lm\in\mathrm{Par}(N)} \HN_T; \label{decN1} \\
\mathcal{M}_N&=& \bigoplus_{T\in\mathcal{T}^S_{\lm},\lm\in\mathrm{Par}(N)}\mathcal{M}_N(T). \label{decN2} 
\end{eqnarray}
Here $\mathcal{M}_N(T)$ is spatially equivalent to $\mathcal{M}_N(T')$ iff $T$ and $T'$ both lie in $\mathcal{T}^S_{\lm}$ (i.e., for the same $\lm$), so that  the decomposition \er{decN1} - \er{decN2} is non-unique (for example,
 Young tableaux different from standard ones might have been chosen in the parametrization). 
Sacrificing the use of true subspaces of $\HN$ in favour of explicit multiplicity, one may 
alternatively give a simultaneous decomposition of $\mathcal{M}_N$ and $U(S_N)$ \emph{up to spatial and unitary equivalence} as
\begin{eqnarray}
\HN&\cong& \bigoplus_{\lm\in\mathrm{Par}(N)} \HN_{T_{\lm}}\ot V_{\lm}, \label{sectors1}\\
\mathcal{M}_N&\cong& \bigoplus_{\lm\in\mathrm{Par}(N)} \mathcal{M}_N(T_{\lm})\ot 1_{V_{\lm}}, \label{sectors2}\\
U(S_N) &\cong& \bigoplus_{\lm\in\mathrm{Par}(N)} 1_{\HN_{T_{\lm}}}\ot U_{\lm},\label{sectors3}
\end{eqnarray}
where this time the labeling is by the partitions of $N$ themselves,  the multiplicity spaces $V_{\lm}$ are  irreducible $S_N$-modules, and $T_{\lm}$ is  an arbitrary choice of a Young tableau  defined on  $F_{\lm}$. 

 For example, the partitions \er{partition} of $N=2$ are $2=2$ and $2=1+1$,  each of which admits only one standard Young tableau, which we denote by $S$ and $A$, respectively.
 With $N_2=N_{1+1}=1$ and hence $V_1\cong V_{1+1}\cong\C$ as vector spaces, this recovers  \er{HSHA};
  the corresponding projections $P_S$ and $P_A$, respectively, are given by
 $P_S=\half (1+U(12))$ and
$P_A=\half (1-U(12))$.
The bosonic states $\ps_S$, i.e., the solutions of $\ps_S\in H^{\ot 2}_S$, or $P_S\ps_S=\ps_S$,
are just the symmetric vectors, whereas the  bosonic states $\ps_A\in H^{\ot 2}_A$ are the antisymmetric ones. These sectors exists for all $N>1$ and they always occur with multiplicity one. However, 
for $N\geq 3$ additional \irrep s of $\mathcal{M}_N$ appear, always with multiplicity greater than one;
 states in such sectors are said to describe \emph{paraparticles} and/or are said to have \emph{parastatistics}.

For example, for $N=3$ one new partition $3=2+1$ occurs, with $N_{2+1}=2$, and hence
\begin{equation}
H^{\ot 3}\cong H^{\ot 3}_S\oplus H^{\ot 3}_A\oplus H^{\ot 3}_P\oplus H^{\ot 3}_{P'},\label{HSHAHP}
\end{equation}
where $H_P^{\ot 3}$ is the image of the projection
$P=\third(1-U(13))(1+U(12))$, and $H_{P'}^{\ot 3}$ is the image of $P'=\third(1-U(12))(1+U(13))$.
The corresponding two classes of parastates $\ps_P$ and $\ps_{P'}$ then by definition 
satisfy $P\ps_P=\ps_P$ and $P'\ps_{P'}=\ps_{P'}$, respectively.
In other words, the \Hs s carrying each of the four sectors are the following closed linear spans:
\begin{eqnarray}
H_S^{\ot 3}&=&\mathrm{span}^-\{\ps_{123} +  \ps_{213}+ \ps_{321}+ 
 \ps_{312}+  \ps_{132}+ \ps_{231}\};\\
H_A^{\ot 3}&=&\mathrm{span}^-\{\ps_{123} -  \ps_{213}- \ps_{321}+ 
 \ps_{312}-  \ps_{132}+ \ps_{231}\};\\
H_P^{\ot 3}&=&\mathrm{span}^-\{\ps_{123}+  \ps_{213}- \ps_{321}- \ps_{312}\};\label{defHP}\\
H_{P'}^{\ot 3}&=&\mathrm{span}^-\{\ps_{123}+  \ps_{321}- \ps_{213}- \ps_{231}\},\label{defHPp}\
\end{eqnarray}
where $\ps_{ijk}\equiv \ps_i\ot\ps_j\ot\ps_k$ and the $\ps_i$ vary over $H$. See  \S\ref{S8} for a realization of $U_P(S_3)$.

Finally, let us note a special feature of the bosonic and fermonic sectors, namely that $S_N$ maps each of the subspaces $H_S$ and $H_A$ into itself; the former is even pointwise invariant under $S_N$, whereas elements of the latter at most pick up a minus sign. This is no longer the case for parastatistics: for example,
for $N=3$ some permutations map $H_P$ into $H_{P'}$, and \emph{vice versa}. This clear from \er{sectors1} -  \er{sectors3}: for $\lm=P$, one has $\dim(V_P)=2$, and choosing a basis $(e_1,e_2)$ of $V_P$ one may identify $H^{\ot 3}_P$ and  $H^{\ot 3}_{P'}$ in \er{decN1} with (say)  
$H^{\ot 3}_P\ot e_1$ and $H^{\ot 3}_P\ot e_2$ in \er{sectors1}, respectively. And analogously for $N>3$, where $\dim(V_{\lm})>1$ for all $\lm\neq S,A$.
\section{Permutation-invariant classical states}\label{S3}
The competing approach to permutation invariance starts from classical (rather than quantal) data.
Let $Q$ be the classical single-particle configuration space, e.g., 
$Q=\R^d$; to avoid irrelevant complications, we assume that $Q$ is a connected and simply connected smooth manifold.
The  associated configuration space of $N$ identical but distinguishable particles is $Q^N$. 
 Depending on the assumption of (in)penetrability of the particles, we may define one of
\begin{eqnarray}
 \breve{Q}_N&=&Q^N/S_{N};\label{Qbar0}\\
 Q_N&=&(Q^N \backslash
\Delta_N)/S_{N} \label{Qbar}
\end{eqnarray}
as the configuration space of $N$ indistinguishable particles.\footnote{Here $\Dl_N$ is the extended diagonal in $Q^N$, i.e., the set of points $(q_1,\ldots, q_N)\in Q^N$ where $q_i=q_j$ for at least one pair $(i,j)$, $i\neq j$ (so that for $Q=\R$ and $N=2$ this is the usual diagonal in $\R^2$).}  Naively, these two choices should lead to exactly the same quantum theory, based on the \Hs\ $L^2( \breve{Q}_N)=L^2(Q_N)$, since $\Dl_N$ is a subset of measure zero for any measure used to define $L^2$ that is locally equivalent to Lebesgue measure. However,  the effect of $\Dl_N$ is noticeable as soon as one represents  physical observables as operators on $L^2$ through any serious quantization procedure, which should be sensitive to both the topological and  the smooth structure of the underlying configuration space \cite{Isham84}.
In the case at hand, $Q_N$ is multiply connected as a topological space but as a manifold it is  smooth, without any singularities. In contrast, 
 $\breve{Q}_N$ is simply connected as a topological space, but in the smooth setting it is a so-called
 \emph{orbifold}.\footnote{See e.g.\ \cite{Moerdijk}. This includes the possibility of a manifold with boundary. This is easily seen for $d=1$ and $N=2$, in which case
$S_2$ acts on $\R^2$ by $(x,y)\mapsto (y,x)$, and hence $\R^2/S_2$ may be identified with all points in $\R^2$ south-east of the diagonal
$\Delta=\{(x,x)\}$, including $\Delta$. A change of coordinates $(x,y)\mapsto(x_+,x_-)$, $x_{\pm}=\half(x\pm y)$ turns this into the upper half plane (including the vertical axis), where $x_+$ is the horizontal axis whilst $x_-$ is the vertical one. The $S_2$-action is then given by $(x_+,x_-)\mapsto (x_+,-x_-)$.} 
In general, there exist various definitions and associated competing (pseudo)differential cacluli of smooth functions on manifolds with singularities like orbifolds, but in the case at hand 
it is natural to define  the classical observables  as the $S_N$-invariant functions
on $T^*(Q^N)$, which choice is just a rephrasing of the notion of indistinguishability in terms of observables rather than states. According to our theory in sections \ref{S6} - \ref{S8}, for $Q=\R^d$ with $d>2$  such functions  are quantized  by the C*-algebra $K(L^2(Q^N))^{S_N}$, whose \irrep s yield the possible superselection sectors or `inequivalent quantizations' of the system.\footnote{Fans of self-adjoint extensions might like to try to obtain fermions and perhaps even paraparticles from different boundary conditions on the Hamiltonian, but this is neither possible nor necessary. Continuing the example in the previous footnote, the $S_2$-invariant free Hamiltonian
$\partial^2/\partial x^2 + \partial^2/\partial y^2$ is transformed to $\half(\partial^2/\partial x_+^2 + \partial^2/\partial x_-^2)$ on the upper half plane. In principle, if this operator is initially defined on $C_c^{\infty}$ functions on the open upper half plane (omitting the  horizontal axis), then it has a one-parameter family of self-adjoint extensions. The single correct one, however, given by the physics,  is the projection of the domain of the Laplacian on $\R^2$ to the $S_2$-invariant (i.e., symmetric) $L^2$-functions in $x_-$. Transformed to the upper half plane, this yields the Neumann  boundary condition $\partial\Psi(x_+,x_-)/\partial x_-)_{|x_-=0}=0$ for (almost) all $x_+$.}
 As we shall see, these coincide with those for the choice $Q_N$,
so that the difference between \er{Qbar0} and \er{Qbar} only shows up in low dimension ($d\leq 2$).
Since \er{Qbar0} has hardly been studied in the literature, practically all of which is concerned with \er{Qbar}, and our main aim is to clarify the literature, we will henceforth stick to the latter (future work should explore  \er{Qbar0} in $d\leq 2$, however).

The main feature of $Q_N$ is that it is multiply connected. Using a variety of different arguments,  the  literature 
\cite{Dowker,Horvathy1,Horvathy2,HorvathyMorandiSudarshan,SudarshanImboImbo,Isham84,LDW,LM1,Morandi1992,SchulmanBook,Souriau1,Souriau2}
concludes that in such a situation one should at least initially define wave-functions on the connected and simply connected \emph{universal covering space} $\til{Q}_N$ of $Q_N$, so that $Q_N\cong \til{Q}_N/\pi_1(Q)$, where $\pi_1(Q)$ is the first (based) homotopy group of $Q_N$, with defining (right-) action on $\til{Q}_N$.
 
  If $Q=\R^d$ with $d>2$ then $\til{Q}_N=\R^{dN}\backslash
\Delta_N$ with $\pi_1(Q_N)=S_N$, but in $d=2$ the space $\R^{2N}\backslash
\Delta_N$ is not simply connected and
one surprisingly has the braid group $\pi_1(Q_N)=B_N$.

In the remainder of this section, $Q_N$ will  denote an arbitrary multiply connected configuration space.\footnote{\label{fncon} In order to avoid confusion with our single-particle configuration space $Q$ we keep the index $N$, which, then, for the moment will not refer to anything.}
The fact that the physical configuration space under consideration is $Q_N$ rather than $\til{Q}_N$  is then typically taken into account by something like the following  postulate,\footnote{\label{multiv}Which some authors actually try to derive \cite{Dowker,Horvathy1,Horvathy2,HorvathyMorandiSudarshan,SudarshanImboImbo,Isham84,LDW,LM1,Morandi1992,SchulmanBook,Souriau1,Souriau2}, sometimes on valid but unnecessarily restrictive grounds, sometimes in a rather dubious way. 
 For example, `taking a wave-function around a closed loop' in a \Hs, as some  apparently are able to do, 
seems hard to follow, as no action of the loop group of $Q$ is defined. Talk of
 `multi-valued wave-functions' is also unrecommended; it is not defined
 for $L^2$-functions, whereas for smooth or continuous functions one should preferably talk about induced
  \rep s realized on spaces of sections of (hermitian) line bundles, see sections \ref{S7} and \ref{S8}.}  which (as discussed below) presupposes the use of scalar wave-functions:\footnote{To their credit, some authors are  quite explicit on this point. For example,  Morandi \cite{Morandi1992} opens his treatment with the qualifier that he `will consider only \emph{scalar} Quantum Mechanics' (emphasis in original).}
\begin{quote}\emph{
The  wave-functions $\til{\psi}$ of a quantum system with multiply connected classical configuration space $Q_N$ can only be of the following kind: given some character $\ch:\pi_1(Q_N)\raw U(1)$, fixed in a given `sector',\footnote{Sometimes called a \emph{superselection sector} or \emph{inequivalent quantization},  terminology we will relate to in \S\ref{S4}.} $\til{\psi}: \til{Q}_N\raw\C$ satisfies the constraint 
   \begin{equation}
  \til{\psi}(\til{q}g)=\ovl{\ch(g)}\til{\psi}(\til{q}), \label{equi1}
\end{equation}for (almost) all $\til{q}\in\til{Q}_N$ and all $g\in\pi_1(Q_N)$. 
  }   \end{quote}
  Note that  if quantum observables are  $\pi_1(Q_N)$-invariant operators  on $L^2(\til{Q}_N)$ with respect to the  unitary \rep\ $U(g)\til{\psi}(\til{q})= \til{\psi}(\til{q}g)$, as one might assume for any decent quantization prescription, then the constraint \er{equi1} is preserved under the action of such operators.
  
 For $Q=\R^d$ in $d>2$, for any $N>1$ this  yields bosons and fermions and nothing else, as well as their realizations in terms of symmetric and anti-symmetric wave-functions, respectively. Indeed, the only characters of $S_N$ are $\ch_S(\pi)=1$ for all $\pi\in S_N$, and $\ch_A(\pi)=1$
 for all even permutations $\pi$ and  $\ch_A(\pi)=-1$ for all odd ones. On the former, bosonic choice the constraint \er{equi1} reads $\til{\psi}(q_{\pi(1)}, \ldots, q_{\pi(N)})=\til{\psi}(q_1,\ldots, q_N)$ for all $\pi\in S_N$, whereas on the latter the right-hand side picks up a minus sign for odd permutations, as befits fermions. 

 Identifying the \Hs\ $H$ of the previous section with $L^2(Q)$, and granting that the (bounded) observables are defined as in \er{defMN}, one may compare the two approaches. Clearly, the present one
 yields exactly the physically desirable sectors,  and excludes parastatistics. 
However, this reasoning  suffers from similar deficiencies as the one discarded near the end of the Introduction. Since all separable \Hs s are isomorphic, particular realizations of states as wave-functions are only meaningful in connection with some action of observables. Granting that $L^2(\til{Q}_N)$ is a useful starting point, provided it is combined with the assumption that quantum observables are $\pi_1(Q_N)$-invariant operators, most arguments leading to \er{equi1} are based on the idea that two unit vectors represent the same state iff they differ by a phase vector. Once again, this idea is wrong precisely in the situation it is supposed to address, where the observables are constrained by permutation invariance. In fact, the only valid conclusion would be that $\til{\psi}$ and $R_g\til{\psi}$ (i.e.\ the function $\til{q}\mapsto \til{\psi}(\til{q}g)$
define the same vector state (in the algebraic sense) on the algebra of $\pi_1(Q_N)$-invariant operators. But this by no means implies \er{equi1}. Therefore,  despite its promising starting point,  the configuration space approach is based on a series of subsequent mathematical assumptions and moves that on closer inspection are somewhat arbitrary. To resolve this, the general interplay between the classical configuration space and the 
quantum observables needs to be clarified first. 
\section{Strict deformation quantization}\label{S4}
The desired interplay between the classical configuration space and the 
quantum observables is provided by strict deformation quantization \cite{book,handbook,Rieffel89,Rieffel94}.
Here $\hbar$ is a genuine real number,\footnote{As opposed to a formal parameter, as in the original idea of deformation quantization due to Berezin \cite{Ber} and Flato et al \cite{BFFS}, which in our opinion is physically less relevant.}  and the quantum system under study is described by a C*-algebra of observables (as usual in algebraic quantum theory \cite{Haag}). Call the latter $\CA_{\hbar}$ - in the examples in this paper,\footnote{However, there are many other examples of  strict deformation quantization in which the different $\CA_{\hbar}$ fail to be isomorphic, starting with Rieffel's original motivating example of the noncommutative torus \cite{Rieffel89}.} all C*-algebras  $\CA_{\hbar}$ will be isomorphic for $\hbar >0$, so we might  as well work with a single C*-algebra $\CA$. Its classical counterpart is a phase space, more precisely, a Poisson manifold $M$.\footnote{This  is a manifold equipped with a Lie bracket $\{\, ,\,\}$ on $\cin(M)$ with the property that for each $f\in\cin(M)$ the map $g\mapsto \{f,g\}$ defines  a derivation of  the commutative algebra structure of $\cin(M)$ given by pointwise multiplication.  Hence this map is given by a vector field $\xi_f$, called
the \emph{Hamiltonian vector field} of $f$ (i.e.\ one has $\xi_fg=\{f,g\}$).
 \emph{Symplectic manifolds} are special instances of Poisson manifolds, characterized by the property that the  Hamiltonian vector fields exhaust the tangent bundle.}  We then say that a noncommutative C*-algebra (of quantum  observables) $\CA$ is a  \emph{strict deformation quantization} of a given 
 Poisson manifold $M$ (with associated \emph{commutative} C*-algebra $C_0(M)$ of classical observables)
 if the following conditions are satisfied:
\begin{itemize}
\item The family $(\CA_0=C_0(M), \CA_{\hbar}=\CA, \hbar\in (0,1])$ forms a continuous field of C*-algebras (cf.\ \cite{Dix}) over $[0,1]$.
\item For each $ \hbar\in (0,1]$, a quantization map $Q_{\hbar}: \cci(M)\raw \CA$ is given such that for each $f\in\cci(M)$ the map $\hbar\mapsto Q_{\hbar}(f)$ is a continuous cross-section of this continuous field of C*-algebras. (These  quantization maps will not play an explicit role in this paper.)
\item 
 For all $f,g\in \cci(M)$ one has 
$\lim_{\hbar\rightarrow 0} 
\left\|\frac{i}{\hbar}[Q_{\hbar}(f),Q_{\hbar}(g)]-Q_{\hbar}(\{f,g\})\right\|_{\CA} =0$.
\end{itemize}
This provides a powerful approach to  quantization, which is both 
physically relevant and mathematically rigorous. For example, Mackey's approach to quantization, which is based on the systematic use of induced \rep s and the associated systems of imprimitivity  \cite{Mackey},
 is a special case of strict deformation quantization, as is Isham's closely related method, based on so-called canonical groups \cite{Isham84}; from our point of view, both effectively use groupoid C*-algebras that play the role of the deformation $\CA$ above \cite{book,Landsman2006,handbook}. In this and other cases, the connection with the physicist's approach  to  quantization is that irreducible representations of $\CA$ yield both the Hilbert space and the commutation relations (the latter simply reflecting the algebraic structure of $\CA$). In particular, the
connection with the  
`inequivalent quantizations' of a phase space $M$ constructed in the physics literature  emerges as 
follows
\cite{landsman1990b,landsman1990c}:
\begin{Dogma}
 The inequivalent quantizations of a phase space $M$ (in the physicist's sense) are given by the inequivalent irreducible representations of  the corresponding algebra of quantum observables $\CA$ (defined mathematically  as a strict deformation quantization of $M$).
\end{Dogma}
We refer to  \cite{book,Landsman2006,handbook}, as well as to the main body of this paper,  for examples. One crucial issue we need to address here is the possible lack of uniqueness of $\CA$, for a given phase space $M$. Let us illustrate this non-uniqueness in the example of spin, which at the same time illustrates the entire procedure in an elementary context.
\section{Spin and its generalizations}\label{S5}
 The classical phase space for a spinning particle is $\R^3$, seen as the dual vector space to the Lie algebra $so(3)$ of the Lie group $SO(3)$, equipped with the so-called Lie--Poisson bracket \cite{Souriau2}. This is given on the standard coordinate functions $(x_1,x_2,x_3)$ on $\R^3$ by $\{x_1,x_2\}=-x_3$ and cyclic permutations thereof; compare this with the usual Poisson bracket on $\R^{2n}$, given on the  coordinate functions $(p_1,\ldots, p_n, q_1, \ldots, q_n)$ by $\{p_i,q_j\}=\delta_{ij}$, $\{p_i,p_j\}=\{q_i,q_j\}=0$. The strict deformation quantization of this phase space is not unique; the 
 group C*-algebra $C^*(G)$ yields one for either $G=SO(3)$ or $G=SU(2)$ \cite{book,Rieffel90}. 
 To see what this means for the inequivalent quantizations of $\R^3$, we recall that for any locally compact group $G$ one has a bijective correspondence between nondegenerate (irreducible) representations $\pi$ of $C^*(G)$ and unitary  (irreducible) representations $U$ of $G$  \cite{Echterhoff},  given by (continuous extension of)
\begin{equation}
\pi(f)=\int_G dx\, f(x) U(x), \: f\in C_c(G).
\end{equation}
Hence  the irreducible representations of $C^*(SU(2))$  are given by
representations $\pi_j$ corresponding  the familiar unitary representations $U_j\equiv \mathcal{D}_j$ of $SU(2)$ on $H_j=\C^{j+1}$, for $j\in\mathbb{N}/2\equiv \{0,1/2, 1, 3/2, 2, \ldots\}$, whereas 
 the irreducible representations of $C^*(SO(3))$ correspond to the unitary representations of $SO(3)$, i.e., to  $\mathcal{D}_j$, but now with $j\in\mathbb{N}=\{0,1,2,\ldots\}$. The physical interpretation of these representations is that they describe immobile particles with  spin $j$; we 
 see that  the choice of  $C^*(SU(2))$ as a strict deformation quantization of $\R^3$ yields all allowed values of quantum spin, whereas $C^*(SO(3))$ only gives half of them. 
 
 The mathematical reason for this is that $SU(2)$ is the unique connected and simply connected Lie group with the given Lie algebra $so(3)$, whereas $SO(3)$ is doubly connected; recall the well-known isomorphism $SO(3)\cong SU(2)/\Z_2$, where 
 $\Z_2=\{1_2,-1_2\}$ is the center of $SU(2)$. At the level of the corresponding group C*-algebras, this isomorphism  becomes
 \begin{equation}
C^*(SO(3))\cong C^*(SU(2))/I_{\Z_2},
\end{equation}
where the ideal $I_{\Z_2}$ in $C^*(SU(2))$ is the norm-closure of the set of $f\in C(G)$ satisfying $f(-x)=-f(x)$. This leads to a reinterpretation of the representation theories of $C^*(SU(2))$ and  $C^*(SO(3))$ just discussed:
the representations of $C^*(SO(3))$ form a subset of those of $C^*(SU(2))$, consisting of the representations of $C^*(SU(2))$ that send the ideal $I_{\Z_2}$ to zero. 

More generally, let $G$ be a connected Lie group with Lie algebra $g$. Then there exists a unique connected and simply connected Lie group $\til{G}$ with the same Lie algebra, and a finite discrete subgroup $Z$ of the center of $\til{G}$, such that $G=\til{G}/Z$.
Representations of $G$ correspond to \rep s of $\til{G}$ that are trivial on $Z$. For the  group C*-algebras we then have
\begin{equation}
C^*(G)\cong C^*(\til{G})/I_Z,
\end{equation}
where the ideal $I_Z$  is the norm-closure of the  $f\in C_c(G)$ satisfying 
\begin{equation}
\sum_{z\in Z} f(xz)=0 
\end{equation}
for all $x\in\til{G}$;
the \rep s of of $\til{G}$ that are trivial on $Z$ are exactly those for which the corresponding \rep\ of $C^*(\til{G})$  map the ideal $I_Z$ to zero.
\section{Intermezzo: Lie groupoids}\label{Intermezzo}
Even more generally, a similar picture holds for Lie groupoids,\footnote{  
Recall that  a \emph{groupoid} is a small category (i.e.\ a category in which the underlying classes are sets) in which each arrow is invertible.  A \emph{Lie groupoid} is a groupoid for which the total space (i.e.\ the set of arrows) $G$
and the base space $G_0$ are manifolds, the source and target maps $s,t: G\raw G_0$ are surjective submersions, and multiplication and inversion  are smooth. Lie groups may be seen as 
 Lie groupoids,  where  $G_0=\{e\}$. See \cite {Mackenzie} for a comprehensive treatment. Each 
manifold $M$ defines the associated  \emph{pair groupoid} with total space $G=M\x M$ and base
$G_0=M$, with  $s(x,y)=y$, $t(x,y)=x$, $(x,y)\inv=(y,x)$,  multplication
$(x,y)(y,z)=(x,z)$, and units $1_x=(x,x)$. Last but not least (in our context), 
the \emph{gauge groupoid} defined by a principal $H$-bundle $P\stackrel{\ta}{\raw}M$ is given by
 $G=P\x_H P$ (which stands for $(P\x P)/H$ with respect to the diagonal $H$-action on $P\x P$), $G_0=M$, $s([p,q])=\ta(q)$, $t([p,q])=\ta(p)$, $[x,y]\inv=[y,x]$, and $[p,q][q,r]=[p,r]$ (here $[p,q][q',r]$
is defined whenever $\ta(q)=\ta(q')$, but to write down the product one picks $q\in\ta\inv(q')$). }
and this generalization is crucial for our story. The essential point is that a Lie groupoid $G$ canonically defines both a Poisson manifold $g^*$ \cite{Weinstein}, \cite[\S III.3.9]{book},  and a C*-algebra $C^*(G)$ 
 \cite[\S II.5]{Connes}, \cite[\S III.3.6]{book}, which turns out to be a strict deformation quantization of $g^*$ \cite{book,Landsman2006,LR}. Here $g^*$ is the dual vector bundle to the Lie algebroid
 $g$ associated to $G$,\footnote{A \emph{Lie algebroid} $A$ over a manifold
$M$ is a vector bundle $A \stackrel{\ta}{\raw} M$ equipped with a vector bundle map
$A \stackrel{\al}{\raw} TM$ (called the \emph{anchor}), as well as
with a Lie bracket $[\, ,\,]$ on the space  $\cin(M,A)$ of smooth
 sections of $A$, satisfying the Leibniz rule
$ [\sg_1,f \sg_2]=f [\sg_1,\sg_2]+ (\al\circ
\sg_1 f) \sg_2$
for all $\sg_1,\sg_2\in\cin(M,A)$ and $f\in\cin(M)$.  
It follows that the map $\sg\mapsto\al\circ\sg: \cin(M,A)\raw \cin(M,TM)$ induced by the anchor is a homomorphism of Lie algebras, where the latter is equipped with the usual commutator of vector fields.
See \cite {CF,Mackenzie}.
For example, the Lie algebroid of a Lie group is just its Lie algebra, the  Lie algebroid  defined by a pair groupoid
$M\x M$ is the tangent bundle $TM$, and  the Lie algebroid of a gauge groupoid $P\x_H P$ is 
$(TP)/H$, where  $\cin(M,TP)^H$, which inherits the commutator from
$\cin(M,TP)$ as the Lie bracket defining the algebroid structure, and is equipped with the projection 
 induced by the push-forward $\ta':TP\raw TM$ of $\ta$.
} and $C^*(G)$ is an appropriate completion of the convolution algebra on $G$ (which for groups is just the usual group algebra). 
For example, for the pair groupioid $G=M\x M$ one has $g^*= T^*M$, the cotangent bundle of $M$ equipped with the usual symplectic and hence Poisson structure, whilst 
$C^*(M\x M)\cong  K(L^2(M))$, i.e.\ the C*-algebra  of compact operators on the $L^2$-space canonically defined by a manifold. 

We will need the following generalization of this example. Let $\ta:P\raw P/H$ be a principal bundle, with
 gauge groupoid $P\times_H P$.
 The associated Poisson manifold is the quotient $(T^*P)/H$, whilst the corresponding C*-algebra
is, primarily and tautologically, $C^*(P\times_H P)$. When $H$ is compact, this algebra is canonically isomorphic to the $H$-invariant compact operators $K(L^2(P))^H$ on $L^2(P)$, and for any (locally compact) $H$ it is isomorphic to $K(L^2(M))\otimes C^*(H)$,
but any explicit isomorphism depends on the choice of a measurable section $s:M\raw P$, which in general cannot be smooth (cf.\  \cite[Thm.\  III.3.7.1]{book} and \S\ref{S7} below).

The  specialization of this example on which our approach to indistinguishable particles relies,
starts from  a connected  manifold $Q_N$, seen as the configuration space of some physical system; here one may have our motivating example \er{Qbar} in mind,\footnote{Recall footnote \ref{fncon} on our notation.} with e.g.\ $Q=\R^d$. This leads to 
the principle bundle defined by $P=\til{Q}_N$, the universal covering space of $Q_N$,
and $H=\pi_1(Q_N)$, the first homotopy group of $Q_N$ (based at some $q_0\in Q_N$), acting on $\til{Q}_N$ in the usual way (from the right), so that the base space is $\til{Q}_N/\pi_1(Q_N)\cong Q_N$.
The associated  gauge groupoid, Poisson manifold, and C*-algebra associated to this bundle are given by
\begin{eqnarray}
G_{Q_N}&=&\til{Q}_N\times_{\pi_1(Q_N)}\til{Q}_N \label{defGQ};\\
g^*_{Q_N}&=&(T^*\til{Q}_N)/\pi_1(Q_N)\cong T^*Q_N;\\
{C^*(G_{Q_N})}&\cong&K(L^2(Q_N))\otimes C^*(\pi_1(Q_N)) \cong K(L^2(\til{Q}_N))^{\pi_1(Q_N)}, \label{pi1f}
\end{eqnarray}
respectively; the last isomorphism holds only if $\pi_1(Q_N)$ is finite (see \S\ref{S6}  for a proof).
 \section{Quantization of multiply connected spaces}\label{S6}
The phase space of a classical system with  configuration space $Q_N$ is the cotangent bundle $T^*Q_N$. 
The simplest way to quantize this using the formalism of the previous section would be to observe that 
as a Poisson manifold, $T^*Q_N$ is the dual $g^*$ to the Lie algebroid $g=TQ_N$ of the pair groupoid $G=Q_N\x Q_N$. Hence the associated groupoid C*-algebra 
\begin{equation}
C^*(Q_N\x Q_N)\cong K(L^2(Q_N)) \label{naive}
\end{equation}
provides a strict deformation quantization of $T^*Q_N$. This is true but incomplete, in a way comparable to quantizing a dual Lie algebra $g^*$ using the C*-algebra $C^*(G)$ of a Lie group $G$ with Lie algebra $g$, where $G$ fails to be simply connected. As we have seen in \S\ref{S5}, using the language of Dogma 1 in \S\ref{S4}, this misses a large number of possible inequivalent quantizations (e.g., for $G=SO(3)$ it  misses all half-integer spins). In the case at hand, using \er{naive} would miss all particle statistics except bosons, which of course is empirically unacceptable. 

The correct quantization procedure copies the one for Lie groups, \emph{mutatis mutandis}.\footnote{The following result is implicit in \cite{CF} and has been made explicit in an email  by Marius Crainic (2011).} 
\begin{Proposition}
For any Lie groupoid $G$, with Lie algebroid $g$, there exists a source-connected and source-simply connected Lie groupoid $\til{G}$ with the same Lie algebroid $g$, unique up to isomorphism. All other 
Lie groupoids    with Lie algebroid $g$ (like $G$) are quotients of $\til{G}$ by some normal subgroupoid $Z$ of $\til{G}$, which is an \'{e}tale bundle of groups (over the base  of $G$).
\end{Proposition}
For $g=TQ_N$, this `universal cover' $\til{G}$ is none other than $G_{Q_N}$ as defined in \er{defGQ}. 
Hence \cite[Thm.\ III.3.12.2, Cor.\ III.3.12.6]{book} (see also \cite[\emph{passim}]{LR}) immediately implies:
\begin{Theorem}\label{t1}
For any connected manifold $Q_N$ with universal cover $\tilde{Q}_N$, the C*-algebra
\begin{equation}
C^*(G_{Q_N})\equiv C^*(\til{Q}_N\times_{\pi_1(Q_N)}\til{Q}_N) \label{AOO}
\end{equation}
is a strict deformation quantization of the Poisson manifold $T^*Q_N$.
\end{Theorem}
Whether or not $\pi_1(Q_N)$ is finite,
from Theorem \ref{t1} and \cite[Cor.\ III.3.7.2]{book}, we next obtain:
\begin{Theorem} \label{t2}
\begin{enumerate}
\item The inequivalent irreducible representations of  $C^*(G_{Q_N})$, and hence (by Dogma 1 in \S\ref{S4}) the inequivalent quantizations of $T^*Q_N$, bijectively correspond to the inequivalent irreducible unitary representations $U_{\chi}$ of $\pi_1(Q_N)$.
\item  Realizing $U_{\ch}$ on a Hilbert space $H_{\chi}$,  the associated
representation $\pi^{\chi}$ of  $C^*(G_{Q_N})$ is naturally realized on the  Hilbert space
$L^2(Q_N,E_{\chi})$
of $L^2$-sections of the vector bundle
 \begin{equation}
E_{\chi}=\til{Q}_N\times_{\pi_1(Q_N)} H_{\chi}  
\end{equation}
 associated to the principal bundle  $\ta: \til{Q}_N\raw Q_N$ by  the \rep\ $U_{\chi}$ on $H_{\chi}$.\footnote{$E_{\chi}$
is defined as the quotient $(\til{Q}_N\x H_{\chi})/\pi_1(Q_N)$ with respect to the action
$h:(\tilde{q},v)\mapsto (\tilde{q}h\inv,U_{\chi}(h)v)\equiv (\tilde{q}h,h\inv v)$, $h\in \pi_1(Q_N)$, $\tilde{q}\in \til{Q}_N$, $v\in H_{\chi}$.
Denoting elements of this set as equivalence classes $[\tilde{q},v]$, the bundle projection $\pi_{\chi}:E_{\chi}\raw Q_N$ is given by $\pi_{\chi}([\tilde{q},v])=\tau(\tilde{q})$
and fiberwise addition is given by  $[\tilde{q},v]+ [\tilde{q}',w]=[\tilde{q},v+h^{-1}w]$, where $h$ is the unique element of $\pi_1(Q_N)$ for which $\tilde{q}'h=\tilde{q}$. See  \cite[\S III.2.1]{book}.}

\end{enumerate}
\end{Theorem}
In principle, these theorems give a complete solution to the problem of quantizing multiply connected configuration spaces, and hence, provided one accepts \er{Qbar}, of the problem of quantizing systems of indistinguishable particles. What remains is to work out the details. 

We start with a proof of the isomorphisms \er{pi1f}, which also gives some insight into what is going on in general. With $G_{Q_N}$ given by \er{defGQ},
a dense set of elements of $C^*(G_{Q_N})$ is given by the space $\cci(G_{Q_N})$ of 
  smooth compactly supported complex-valued  functions on $G_{Q_N}$.
  
If  $\pi_1(Q_N)$ is  finite, as in the case $\pi_1(Q_N)=S_N$, a smooth element $\til{A}\in\cci(G_{Q_N})$ of $C^*(G_{Q_N})$
bijectively    corresponds to a $\cci$ function $A$ on
   $\til{Q}_N\x\til{Q}_N$ satisfying
$A(\til{q}h,\til{q}'h)=A(\til{q},\til{q}')$ for all $h\in \pi_1(Q_N)$
by $A(\til{q},\til{q}')=\til{A}([\til{q},\til{q}'])$, where
 $[\til{q},\til{q}']$ denotes the equivalence class of $(\til{q},\til{q}')\in \til{Q}_N\x\til{Q}_N$ under the diagonal action of $\pi_1(Q_N)$. We write $A\in\cci(\til{Q}_N\x\til{Q}_N)^{\pi_1(Q_N)}$; for \er{Qbar} this just means that $A$ is a permutation-invariant kernel.
We  equip  $\til{Q}_N$  with some measure $d\til{q}$  that  is locally equivalent  to the
 Lebesgue measure, as well as  $\pi_1(Q_N)$-invariant under the `regular' action $R$  of $\pi_1(Q_N)$ on functions on $\til{Q}_N$, given by
$R_h\til{\psi}(\til{q})=\til{\psi}(\til{q}h)$. In that case,  one also has a measure $dq$ on $Q_N$ that is locally equivalent  to the Lebesgue measure, so that  the measures  $d\til{q}$ and $dq$  on   $\til{Q}_N$ and  $Q_N$, respectively, are related by
\begin{equation}
\int_{\til{Q}_N}d\til{q}\,  f(\til{q})=\frac{1}{|\pi_1(Q_N)|}\sum_{h\in\pi_1(Q_N)}\int_{Q_N} dq\, f(s(q)h),\label{intqq}
\end{equation}
where $f\in C_c(\til{Q}_N)$, $|\pi_1(Q_N)|$ is  the number of elements of $\pi_1(Q_N)$, and $s:Q_N\raw\til{Q_N}$ is any (measurable) cross-section of $\ta: \til{Q_N}\raw Q_N$.
We  may then define a Hilbert space  $L^2(\til{Q}_N)$ with respect to $d\til{q}$, on which 
 $\cci(\til{Q}_N\x\til{Q}_N)^{\pi_1(Q_N)}$ acts faithfully  by
\begin{equation}
A\til{\psi}(\til{q})=\int_{\til{Q}_N}d\til{q}'\, A(\til{q},\til{q}')\til{\psi}(\til{q}'). \label{kernel}
\end{equation}
 The product of two such operators is given by the multiplication of the kernels, 
on $\til{Q}_N$,
and involution is as expected, too, namely by `hermitian conjugation', i.e.,
$A^*(\til{q},\til{q}')=\overline{A(\til{q'},\til{q})}$.
The norm-closure of $\cci(\til{Q}_N\x\til{Q}_N)^{\pi_1(Q_N)}$, represented as operators on
$L^2(\til{Q}_N)$ by \er{kernel},  is then given by  $K(L^2(\til{Q}_N))^{\pi_1(Q_N)}$. Hence if $\pi_1(Q_N)$ is finite we have the last isomorphism in \er{pi1f}.\footnote{For experts: the above procedure really proves the isomorphism
$C^*_r(G_{Q_N})\cong K(L^2(\til{Q}_N))^{\pi_1(Q_N)}$, but for finite $\pi_1(Q_N)$ the groupoid $G_{Q_N}$ is amenable, so that $C^*_r(G_{Q_N})\cong C^*(G_{Q_N})$.} The first isomorphism in \er{pi1f}, which always holds, follows from  \cite[Thm.\ III.3.7.1]{book}.
However, whereas the second isomorphism  can already be implemented at the smooth level,  the first is  only true upon completion of the smooth kernels in question into C*-algebras. 

In connection with Theorem \ref{t2}, there are various ways of realizing the Hilbert space  $L^2(Q_N,E_{\chi})$, which enable us to relate our approach to the physics literature. 
The first realization corresponds to having constrained wave-functions defined on the covering space $\til{Q}_N$; for example, the usual description of bosonic or fermonic wave-functions is of this sort. The second uses unconstrained wave-functions on the actual configuration space $Q_N$.\footnote{Such function are often confusingly called `multi-valued' by physicists; see also footnote \ref{multiv}.}
\smallskip

 \textbf{1.} The space $\Gamma(Q_N,E_{\ch})$ of \emph{smooth} cross-sections of $E_{\chi}$ may be given by the smooth maps
$\tilde{\psi}: \til{Q}_N\raw H_{\chi}$
 satisfying
the equivariance condition (`constraint')
\begin{equation}
\tilde{\psi}(\tilde{q}h)=U_{\ch}(h\inv)\tilde{\psi}(\tilde{q}), \label{equi}
\end{equation}
for all  $h\in \pi_1(Q_N)$, $\tilde{q}\in \til{Q}_N$. 
To define a Hilbert space,
note that for any $\tilde{\psi},\tilde{\phv}\in\Gamma(Q_N,E_{\ch})$ the function $\til{q}\mapsto (\tilde{\psi}(\til{q}),\tilde{\phv}(\til{q}))_{H_{\chi}}$
is invariant under $\pi_1(Q_N)$ by \er{equi} and unitarity of $U_{\ch}$, and hence defines a function on $Q_N$. Hence   we may define a sesquilinear form on  $\Gamma(Q_N,E_{\ch})$ by
\begin{equation}
(\tilde{\psi},\tilde{\phv})=\int_{Q_N} dq\,  (\tilde{\psi}(\til{q}),\tilde{\phv}(\til{q}))_{H_{\chi}},
\end{equation}
where $q=\tau(\til{q})$, and $dq$ is related to $d\til{q}$ by \er{intqq} (omitting the factor $1/|\pi_1(Q_N)|$ if $\pi_1(Q_N)$ is infinite).  The Hilbert space 
\begin{equation}
H^{\chi}=L^2(\til{Q}_N, H_{\chi})^{\pi_1(Q_N)}, \label{defHchieq}
\end{equation}
then, is defined as the usual $L^2$-completion of the space of all
$\til{\psi}\in \Gamma(Q_N,E_{\ch})$ for which $(\til{\psi},\til{\psi})<\infty$. The \irrep\ $\pi^{\chi}(C^*(G_{Q_N}))$ 
is then given on elements $\til{A}$ of the dense subspace $\cci(G_{Q_N})$ of $C^*(G_{Q_N})$ by the expression
\begin{equation}
\pi^{\chi}(\til{A})\psi(\til{q})=\int_{\til{Q}_N}d\til{q}'\, \til{A}([\til{q},\til{q}'])\ps(\til{q}');
\end{equation}
in fact, any  $\pi_1(Q_N)$-invariant operator on $L^2(\til{Q}_N)$ acts on $H^{\chi}$ in this way (by ignoring $H_{\ch}$).

If $\pi_1(Q_N)$ is finite, then two simplifications occur. Firstly, $H_{\ch}$ is finite dimensional, and secondly each Hilbert space $H^{\ch}$ may be regarded as a subspace of  $L^2(\til{Q}_N)$; the above action of $C^*(G_{Q_N})$ on $H^{\ch}$ is then simply given by restriction of its action on $L^2(\til{Q}_N)$.  In that case one may equivalently realize this \irrep\ in terms of the right-hand side of \er{pi1f}, in which case 
the action of $\pi^{\chi}(A)$ 
on $H^{\ch}$ as defined in \er{defHchieq} is given by
\begin{equation}
\pi^{\chi}(A)\psi(\til{q})=\int_{\til{Q}_N}d\til{q}'\, A(\til{q},\til{q}')\ps(\til{q}').  \label{piA}
\end{equation}
This is true as it stands if $A\in\cci(\til{Q}_N\x\til{Q}_N)^{\pi_1(Q_N)}$, i.e., $A(\til{q}h,\til{q}'h)=A(\til{q},\til{q}')$ for all $h\in \pi_1(Q_N)$, and may be 
extended to general $A\in  K(L^2(\til{Q}_N))^{\pi_1(Q_N)}$ by norm continuity, and even to $B(L^2(\til{Q}_N))^{\pi_1(Q_N)}$ by strong or weak  continuity. See also \cite[Cor.\ III.3.7.2]{book}. 
\smallskip

\textbf{2.} Note that the elements of the Hilbert space $L^2(\til{Q}_N, H_{\chi})^{\pi_1(Q_N)}$ are typically (equivalence classes of) \emph{discontinuous} cross-sections of $E_{\ch}$. However,
possibly discontinuous cross-sections
 may simply be given directly as  functions  
  $\psi: Q_N \raw H_{\ch}$, with inner product
  \begin{equation}
(\psi,\phv)=\int_{Q_N} dq\,  (\psi(q),\phv(q))_{H_{\chi}}.
\end{equation}
This specific realization of $L^2(Q_N,E_{\chi})$ will be denoted by $L^2(Q_N)\otimes H_{\ch}$. 
Of course, in case that $H_{\ch}=\C$, one has the further simplification $L^2(Q_N)\otimes H_{\ch}\cong L^2(Q_N)$.
 Let us also note that, since at the Hilbert space level one is working in a measurable (as opposed to a continuous or smooth) context, in the above formulae one may replace the configuration space $Q_N$ by a fundamental domain $\Delta$ for $\pi_1(Q_N)$ in $\til{Q}_N$, so that  $L^2(Q_N)\otimes H_{\ch}$ is replaced by  $L^2(\Delta)\otimes H_{\ch}$.  E.g., in the example in \S\ref{S7} one has
 $Q_N=S^1$, $\til{Q}_N=\R$, $\pi_1(Q_N)=\Z$, $\Delta=[0,1)$.
\smallskip

The equivalent descriptions \textbf{1} and  \textbf{2} 
 may be related once a (typically discontinuous) cross-section $\sg:Q_N\raw \til{Q}_N$ of the projection 
$\tau: \til{Q}_N\raw Q_N$ has been chosen (i.e., $\tau\circ\sg=\mathrm{id}_{Q_N}$), in which case
 $\psi(q)=\tilde{\psi}(\sg(q))$. 
We formalize this in terms of a unitary operator
\begin{eqnarray}
U&:& L^2(\til{Q}_N, H_{\chi})^{\pi_1(Q_N)}\raw L^2(Q_N)\otimes H_{\ch}\label{Usg0}\\
{U\til{\psi}(q)} &=& \tilde{\psi}(\sg(q));\\
{U\inv\psi(\til{q})} &=& U_{\ch}(h)\psi(q),\label{Usg}
\end{eqnarray}
where $q=\tau(\til{q})$, and $h$  is the unique element of $\pi_1(Q_N)$ for which $\tilde{q}h=\sg(q)$.
The  action $\pi^{\ch}_{\sg}(A)=U\pi^{\ch}(A)U\inv$  on the `unconstrained' wave-functions in $L^2(Q_N)\otimes H_{\ch}$ now follows from \er{piA} - \er{Usg}:
 if $A$ is  a $\pi_1(Q_N)$-invariant kernel  on $L^2(\til{Q}_N)$,  
then using \er{intqq} we obtain
\begin{equation}
\pi^{\ch}_{\sg}(A)\psi(q)=\sum_{h\in\pi_1(Q_N)}\int_{Q_N} dq'\, A(\sg(q),\sg(q')h)U_{\chi}(h)\psi(q'). \label{pichi}
\end{equation}
\section{Particle on a circle}\label{S7}
This example has already been treated from the same conceptual point of view as in  the present paper  \cite{landsman1990b,landsman1990c}, but the mathematical language  was a little different (i.e., transformation group C*-algebras as opposed to groupoid C*-algebras), so we briefly return to it. Let
$Q_N=S^1$ (i.e., the circle), so that $\til{Q}_N=\R$ and $\pi_1(Q_N)=\Z$. In that case one has
\begin{equation}
C^*(G_{Q_N})=C^*(\R\times_{\Z}\R) \cong C^*(\R,\R/\Z),
\end{equation}
 the transformation group C*-algebra $C^*(G,G/H)$ with $G=\R$ and $H=\Z$; see  \cite[Cor.\ III.3.7.5]{book}. The label $\chi$ for unitary \irrep s of $\pi_1(Q_N)$ is now played by the famous $\theta$-angle, i.e.,
one has $U_{\theta}(n)=\exp(in\theta)$, $n\in \Z$, on $H_{\theta}=\C$, where $\theta\in [0,2\pi)$.

 The superselection sectors/inequivalent quantizations, then, may be  realized as follows. 
\begin{enumerate}
\item The realization of  $H^{\theta}$ as $L^2(\R,H_{\theta})^{\Z}$ now
consists of all measurable functions $\tilde{\psi}:\R\raw\C$ satisfying the constraint $\ps(x+n)=\exp(in\theta)\til{\psi}(x)$ for all $n\in Z$ and $\int_0^1 dx\, |\ps(x)|^2<\infty$.\footnote{$H^{\theta}$ is not a subspace of $L^2(\R)$, since
functions satisfying $\ps(x+n)=\exp(in\theta)\til{\psi}(x)$ are not in $L^2(\R)$.}
\item The realization  of $H^{\theta}$ as $L^2(\Delta)\otimes H_{\theta}$, on the other hand, is  $L^2([0,1))=L^2([0,1])$.
\end{enumerate}
\begin{itemize}
\item 
\emph{Position}. A global position coordinate on the circle does not exist and is to be replaced by the space of all continuous (or smooth) functions $f$ on the circle \cite{Isham84}. As is the case for any (infinite) configuration space, such functions do not correspond to elements of the algebra of observables $C^*(\R\times_{\Z}\R)$, but they may be treated in the above way if we extend $f:S^1\raw\C$ to a periodic function $\til{f}:\R\raw\C$ and let $\til{f}$ act on $L^2(\R)$ as a multiplication operator. Indeed, such $\til{f}$ is a $\Z$-invariant operator with respect to the 
regular \rep\  $R$ of $\Z$ on $L^2(\R)$ given by $R_n\ps(x)=\ps(x+n)$,  and hence $\til{f}$ may be seen as an element of a suitable completion of the algebra of observables.\footnote{Alternatively, if we 
extend the kernels $A$ on $L^2(\til{Q}_N)$ to distributions, one has $A(x,y)=\til{f}(x)\delta(x-y)$.}
On $L^2([0,1])$, the expression  \er{pichi} then simply yields $\pi^{\theta}(f)\psi(x)=f(x)\psi(x)$, that is, functions of position are represented as multiplication operators, as usual in the Schr\"{o}dinger \rep.
\item
\emph{Momentum}.
 Explicit $\theta$-dependence appears in the momentum operator (and thence in the Hamiltonian), but in a subtle way, namely through its domain. To see this from the above description, note that $\til{p}=-id/dx$ is essentially self-adjoint on the domain $\Gamma(S^1,E_{\theta})\subset L^2(\R,H_{\theta})^{\Z}$. Passing to $L^2([0,1])$ as in \er{Usg}, the image $\hat{\rho}_{\theta}$ of this domain consists of all
 $\ps\in\cin([0,1])$ satisfying the boundary condition $\ps(1)=\exp(i\theta)\psi(0)$, on which domain the operator $\til{p}$ acts as the usual momentum operator $p=-id/dx$, which is essentially self-adjoint. Hence it has a unique self-adjoint extension $p_{\theta}$, where again the explicit $\theta$-dependence lies in its domain rather than in its `formula'.  
 
 As explained in \cite{landsman1990c}, one may transfer the $\theta$-dependence from the domain to the `formula' by a further unitary transformation on $L^2([0,1])$, which yields  a domain $\hat{\rho}_0$ of essential self-adjointness (consisting of all smooth periodic wave-functions $\psi$) and a `formula' $p_{\theta}=-id/dx +\theta/2\pi$. Another way to get rid of the $\theta$-dependence of the domain of the momentum operator is to consider the one-parameter unitary group $a\mapsto U_a=\exp(ipa)$ on $L^2([0,1])$ generated by $p$, which is defined on the entire Hilbert space and explicitly contains $\theta$, viz.\
 $U_a\ps(x)=\exp(ia\theta/2\pi)\ps(x+a\: \mathrm{mod}\: 1)$.
 \end{itemize}
\section{Indistinguishable particles}\label{S8}
 Our main example is the configuration space \er{Qbar} with $Q=\R^3$, or $Q=((\R^3)^N-\Delta_N)/S_N$,
describing $N$ indistinguishable particles in $\R^3$. The space
$\mathring{\R}^{3N} \equiv(\R^3)^N-\Delta_N$
 is connected and simply connected (unlike its counterpart in $d=2$), so that $\til{Q}_N=\mathring{\R}^{3N}$ and hence
\begin{equation}
\pi_1(Q_N)=S_N.
\end{equation}

$\mathbf{N=1}$. Here $\til{Q}_1=Q_1=\R^3$ and $\pi_1(Q_1)=\{e\}$, so that
 $C^*(G_{Q_1})=K(L^2(\R^3))$,  which has a  unique \irrep\ on $L^2(\R^3)$.\footnote{This is essentially Rieffel's version of the Stone--von Neumann uniqueness theorem \cite{Rieffel72}.}

\medskip

$\mathbf{N=2}$. See also \S\ref{S1}. 
This time, $\pi_1(Q_2)=S_2=\Z_2=\{e,(12)\}$, which has two \irrep s: firstly, $U_B(\pi)=1$
 for both $\pi\in S_2$, and secondly, $U_F(e)=1$, $U_F(12)=-1$, each realized on $H_{\lm}=\C$.
 Hence with $q=(x,y,z)\in\R^3$, eq.\ \er{equi} yields
 \begin{eqnarray}
H^{\ot 2}_B&=& \{\ps\in L^2(\R^3)^{\otimes 2} \mid \ps(q_2,q_1)=\ps(q_1,q_2)\}; \label{N21}\\
H^{\ot 2}_F&=& \{\ps\in L^2(\R^3)^{\otimes 2}\mid \ps(q_2,q_1)=-\ps(q_1,q_2)\}. \label{N22}
\end{eqnarray}
Here $L^2(\R^3)^{\otimes 2}\equiv  L^2(\R^3)\otimes  L^2(\R^3)\cong L^2(\R^6)$.
The algebra 
\begin{equation}
C^*(G_{Q_2})= K(L^2(\R^3)^{\otimes 2})^{S_2} 
\end{equation}
consists of all $S_2$-invariant compact operators on $L^2(\R^6)$, acting on $H^{\ot 2}_B$ or $H^{\ot 2}_F$ in the same way as they do on  $L^2(\R^6)$; cf.\ \er{piA}, noting that (as always) the constraints in 
\er{N21} and \er{N22} are preserved due to the $S_2$-invariance of $A\in C^*(G_{Q_2})$.
This recovers exactly the description of bosons and fermions in \S\ref{S1} and \S\ref{S2}. 
 However, as a warm-up to the next case $N=3$, let us give an alternative realization of  $\pi^F(C^*(G_{Q_2}))$, cf.\ Theorem \ref{t2}. Take two isospin doublet bosons  (i.e., transforming under the defining spin-$\half$ \rep\ of $SU(2)$ on $\C^2$).
With 
\begin{equation}
H^{(2)}=(L^2(\R^3)\otimes\C^2)^{\otimes 2},
\end{equation}
and using indices $a_1,a_2=1,2$,  the 
Hilbert space of these bosons is
\begin{equation}
H_B^{(2)}=\{\ps\in  H^{(2)}
\mid (\ps_{a_2a_1}(q_2,q_1)=\ps_{a_1a_2}(q_1,q_2)\},
\end{equation}
with corresponding
projection $P_B^{(2)}:H^{(2)} \raw H_B^{(2)}$ given by
\begin{equation}
P_B^{(2)}\ps_{a_1a_2}(q_1,q_2)=\half(\ps_{a_2a_1}(q_2,q_1)+\ps_{a_1a_2}(q_1,q_2)).
\end{equation}
Subsequently,  define a partial isometry $W: H^{(2)} \raw L^2(\R^3)^{\otimes 2}$ by 
\begin{equation}
W\ps\equiv \ps_0(q_1,q_2)=\frac{1}{\sqrt{2}}(\ps_{12}(q_1,q_2)-\ps_{21}(q_1,q_2));
\end{equation}
physically, this singles out an isospin singlet Hilbert subspace $H^{(0)}=P_0 H^{(2)}$ within $ H^{(2)}$, where $P_0$ is the projection $W^*W$.
This singlet subspace may be constrained to the bosonic sector by passing to $H^{(0)}_B=P_0P^{(2)}_B H^{(2)}$; note that $P_0$ and $P^{(2)}_B$ commute. 
Now extend the defining \rep\ of $C^*(G_{Q_2})$ on  $L^2(\R^3)^{\otimes 2}$ to $H^{(2)}$ by doing nothing on the indices $a_1,a_2$ (that is, isospin  is deemed to be unobservable). This extended \rep\ commutes with $P_0$ and with $P^{(2)}_B$, and hence is well defined on $H^{(0)}_B\subset H^{(2)}$.
Let us call it $\pi^{(0)}_B$.
\begin{Proposition}
The \rep s $\pi^{(0)}_B(C^*(G_{Q_2}))$ on $H^{(0)}_B$ and $\pi^F(C^*(G_{Q_2}))$ on $H^F$ are unitarily equivalent. 
\end{Proposition}
This is immediate from the fact that
$\ps_0(q_2,q_1)=-\ps_0(q_1,q_2)$.
 In other words, two \emph{fermions} without internal degrees of freedom are equivalent to the singlet state of two \emph{bosons} with an isospin degrees of freedom, at least if the observables are isospin-blind. Similarly,
two \emph{bosons} without internal degrees of freedom are equivalent to the singlet state of two \emph{fermions} with isospin, and two fermions without internal degrees of freedom are equivalent to the isospin triplet state of two fermions.\footnote{This corresponds to the Schur decomposition of $(\C^2)^{\otimes 2}$ under the commuting actions of $S_2$ and $SU(2)$.}

\medskip

$\mathbf{N=3}$.  Here $\pi_1(Q_3)=S_3$, which besides the irreducible boson and fermion \rep s on $\C$ has an irreducible \emph{parafermion} \rep\ $U_{P}$ on $H_P=\C^2$, cf.\ \S\ref{S2}. This \rep\ is most easily obtained explicitly by reducing the natural action of $S_3$ on $\C^3$. Define an orthonormal basis of the latter by
\begin{equation}
e_0=\frac{1}{\sqrt{3}}
\left(
\begin{array}{c}
1  \\
 1\\ 1
\end{array}
\right); \:\:  e_1=\frac{1}{\sqrt{2}}
\left(
\begin{array}{c}
0  \\
 1\\ -1
\end{array}
\right); \:\:
e_2=\frac{1}{\sqrt{6}}
\left(
\begin{array}{c}
-2  \\
 1\\ 1
\end{array}
\right).
\end{equation}
It follows that $\C\cdot e_0$ carries the trivial \rep\ of $S_3$, whereas the linear span of $e_1$ and $e_2$ carries a two-dimensional \irrep\ $U_{P}$, given on the generators by
\begin{equation}
U_{P}(12) =\half \left(
\begin{array}{cc}
1  &-\sqrt{3}      \\
-\sqrt{3}   &  -1
\end{array}
\right);\:
U_{P}(13) =\half \left(
\begin{array}{cc}
1  &\sqrt{3}      \\
\sqrt{3}   &  -1
\end{array}
\right);\:
U_{P}(23) = \left(
\begin{array}{cc}
-1  &0  \\
0   &  1
\end{array}
\right).\label{defUP}
\end{equation}
We already gave realizations of the Hilbert space $H^{\ot 3}_{P}$ of three parafermions in \er{defHP} and  \er{defHPp} in \S\ref{S2}, where it emerged as a subspace of $L^2(\R^3)^{\ot 3}\cong L^2(\R^9)$. An equivalent realization $H^P\equiv \til{H}^{\ot 3}_{P}$ may  be given on the basis of  \er{equi}, according to which $H^P$  is 
 the subspace of
$L^2(\R^3)^{\otimes 3}\otimes \C^2 \cong L^2(\R^9)\otimes\C^2$ that consists of doublet wave-functions $\ps_i$, $i=1,2$, satisfying
\begin{equation}
\ps_i(q_{\pi(1)},q_{\pi(2)}, q_{\pi(3)})=\sum_{j=1}^2 U_{ij}(\pi)\ps_j(q_1,q_2,q_3)
\end{equation}
for any permutation $\pi\in S_3$,
where $U\equiv U_P$, cf.\ \er{defUP}.
In other words, the `parafermionic' wave-functions in this particular realization of $H^{\ot 3}_{P}$ are constrained by the conditions
\begin{eqnarray}
\ps_1(q_2,q_1,q_3)&=&\half \ps_1(q_1,q_2,q_3)-\half \sqrt{3}\, \ps_2 (q_1,q_2,q_3);\\
\ps_2(q_2,q_1,q_3)&=&-\half \sqrt{3}\, \ps_1 (q_1,q_2,q_3) -\half \ps_2(q_1,q_2,q_3);\\
\ps_1(q_3,q_2,q_1)&=&\half \ps_1(q_1,q_2,q_3)+\half \sqrt{3}\, \ps_2 (q_1,q_2,q_3);\\
\ps_2(q_3,q_2,q_1)&=&\half \sqrt{3}\, \ps_1 (q_1,q_2,q_3) -\half \ps_2(q_1,q_2,q_3);\\
\ps_1(q_1,q_3,q_2)&=& - \ps_1(q_1,q_2,q_3); \\
\ps_2(q_1,q_3,q_2)&=&  \ps_2(q_1,q_2,q_3). 
\end{eqnarray}
The algebra of observables of three indistinguishable particles without internal d.o.f., i.e.,
\begin{equation}
C^*(G_{Q_3})= K(L^2(\R^3)^{\otimes 3})^{S_3}
\end{equation}
 then acts on $H^P\subset L^2(\R^3)^{\otimes 3}\otimes \C^2$ as in \er{pichi}, 
 identifying $A\in C^*(G_{Q_3})$ with $A\ot 1_2$ (so that $A$ ignores the internal d.o.f.\  $\C^2$). This \rep\  $\pi^{P}$ is irreducible by Theorem \ref{t2}.

The question now arises where these parafermions are to be found in Nature, or, indeed, whether this question is even well defined!
For we may carry out a similar trick as for $N=2$, and replace parafermions without (further) degrees of freedom by either bosons or fermions. We discuss the former and leave the explicit description of alternatives to the reader.

We proceed as for $N=2$, \emph{mutatis mutandis}. We have a Hilbert space
\begin{equation}
H^{(3)}=(L^2(\R^3)\otimes\C^2)^{\otimes 3},
\end{equation}
of three distinguishable isospin doublets, containing the Hilbert space $H^{(3)}_B$ of three bosonic  isospin doublets as a subspace, that is,
\begin{equation}
H_B^{(3)}=\{\ps\in H^{(3)} \mid \ps_{a_{\pi(1)}a_{\pi(2)}a_{\pi(3)}}(q_{\pi(1)},q_{\pi(2)},q_{\pi(3)})=\ps_{a_1a_2a_3}(q_1,q_2,q_3)\}\, \forall \pi\in S_3.
\end{equation}
The corresponding projection, denoted by $P_B^{(3)}:H^{(3)} \raw H_B^{(3)}$, will not be written down explicitly. 
Define an $SU(2)$ doublet $(\ps_1,\ps_2)$ within the space $H^{(3)}$ through  a partial isometry $W: H^{(3)} \raw L^2(\R^3)^{\otimes 3}\otimes\C^2$,
given by 
\begin{eqnarray}
W\ps_1(q_1,q_2,q_3)&=& \frac{1}{\sqrt{2}}(\ps_{121}(q_1,q_2,q_3)-\ps_{112}(q_1,q_2,q_3)); \\
W\ps_2(q_1,q_2,q_3)&=& \frac{1}{\sqrt{6}}(-2\ps_{211}(q_1,q_2,q_3)+\ps_{121}(q_1,q_2,q_3)+\ps_{112}(q_1,q_2,q_3)).
\end{eqnarray}
Defining a projection $P_2=W^*W$ on $H^{(3)}$, the Hilbert space $H^{(3)}$ contains a closed subspace 
$H^{(2)}_B=P_2P^{(3)}_B H^{(3)}$, which is stable under 
 the natural \rep\ of $C^*(G_{Q_3})$ (since $P_2$ and $P^{(3)}_B$ commute). We call this \rep\ $\pi^{(2)}_B$. An easy calculation then establishes:
 \begin{Proposition}
The \rep s $\pi^{(2)}_B(C^*(G_{Q_3}))$ on $H^{(2)}_B$ and $\pi^{P}(C^*(G_{Q_3}))$ on $H^{P}$ (as  defined by Theorem \ref{t2})  are unitarily equivalent. 
\end{Proposition}
 In other words, three \emph{parafermions} without internal degrees of freedom are quivalent to an isospin doublet formed by three identical \emph{bosonic} isospin doublets,\footnote{This corresponds to the Schur decomposition of $(\C^2)^{\otimes 3}$ under the commuting actions of $S_3$ and $SU(2)$. In this decomposition, the spin 3/2  \rep\ of $SU(2)$ couples to the bosonic \rep\ of $S_3$, whilst the  spin-$\half$ \rep\ of $SU(2)$ couples to the parafermionic \rep\ of $S_3$. 
 } at least if the observables are isospin-blind. And many other realizations of parafermions in terms of fermions or bosons with an internal degree of freedom can be constructed in a similar way. 

\medskip

$\mathbf{N>3}$. 
The above construction may be generalized to any $N>3$. There will now be many parafermionic \rep s $U_{\chi}$ of $S_N$
 (given by a Young tableau), but each of these induces an \irrep\ of the algebra of observables
 \begin{equation}
C^*(G_{Q_N})= K(L^2(\R^3)^{\otimes N})^{S_N}
\end{equation}
 that is unitarily equivalent to a \rep\ on some $SU(n)$ multiplet of bosons with an internal degree of freedom.\footnote{The appropriate multiplet is exactly the one coupled to $U_{\chi}$ in the Schur reduction of $(\C^n)^{\otimes N}$ with respect to the natural and commuting actions of $S_N$ and $SU(n)$ \cite{GoodmanWallach}.
 }
\medskip

The moral of this story  is that one cannot tell from glancing at some Hilbert space whether the world consists of fermions or bosons or parafermions; what matters is the Hilbert space \emph{as a carrier of some (irreducible) \rep\ of the algebra of observables}. From that perspective we already see for $N=2$ that being bosonic or fermionic is not an invariant property of such \rep s,  since one may freely choose between fermions/bosons without internal degrees of freedom and bosons/fermions with those. 
See also \S\ref{S9} no.\ 1 below.
\section{Discussion}\label{S9}
In this final section, we discuss various loose ends and question related to our work. 
\smallskip

\noindent \textbf{1.} The \emph{abelian} (or `scalar') representations of $\pi_1(Q_N)$  play no special role in our approach. In superselection theory one may impose physical selection criterion in order to restrict attention to `physically interesting' sectors. 
Such criteria (which, for example, would have the goal of excluding parastatistics) should be formulated with reference to some algebra of observables. Such issues cannot be settled at the level of quantum mechanics and instead require quantum field theory. Indeed, in (algebraic) quantum field theories with local charges, 
parastatistics can always be removed in terms of either bose- or fermi-statistics, in somewhat similar vein to our \S\ref{S8},  see  \cite{BHS,DHR1,DHR2, DrHR}. For (nonlocal) charges in gauge theories there are no rigorous results, but a similar goal played a role in the road to quantum chromodynamics \cite{French,GellMann}.
\smallskip

\noindent 
 \textbf{2.}  D\"{u}rr et al \cite{Bohmians} have recently argued that in Bohmian mechanics only abelian \rep s of 
 $\pi_1(Q_N)$ can occur, which would imply
  the bose-fermi alternative.  This restriction originates in the requirement that the velocity field $v=\hbar\, \mathrm{Im}(\nabla\ps/\ps)$ be well defined, which in turn leads to the constraint \er{equi1}. This conclusion seems correct, but it is indeed peculiar to  Bohmian mechanics (practitioners of ordinary quantum mechanics in \Hs\ would get a heart attack if they saw the above expression for $v$!).
\smallskip

\noindent 
\textbf{3.} \emph{Does quantization commute with reduction?} 

\noindent  At PSA2010 (cf.\  \cite{Caulton}),  Caulton asked if the following procedures yield the same result:
\begin{itemize}
\item Quantization after reduction:
\begin{enumerate}
\item first impose the identity of the $N$ particles at the classical level;
\item then quantize.
\end{enumerate}
\item Reduction after quantization:
\begin{enumerate}
\item first quantize a system of \emph{a priori} distinguishable particles;
\item then impose the identity of the $N$ particles at the quantum level.
\end{enumerate}
\end{itemize}
 Identifying the second procedure with the Messiah--Greenberg approach of our \S\ref{S2} (which incorporates parastatistics), and associating the first procedure with the quantization of the configuration space \er{Qbar},  the available literature so far (which concludes that the latter approach excludes parastatistics), 
suggests that the answer is \emph{no}. However, 
 our answer  is \emph{yes}, even on the same identifications, since, as we have argued, both approaches lead to the same sectors. The discrepancy originates in the  different quantization procedure we use.
\smallskip

\noindent 
\textbf{4.} Our choice  \er{AOO} as a quantum algebra of observables seems more straightforward than the one by Morchio and Strocchi \cite{morchio2007quantum}. The \emph{regular}  irreducible representations of their algebra of observables 
 bijectively correspond to the irreducible unitary representations of $\pi_1(Q_N)$, and hence also to our inequivalent quantizations. 
 The situation is similar to the use of the C*-algebra of the Weyl form of the canonical commutation relations on $T^*\R^3$ versus the use of the compact operators on $L^2(\R^3)$: the latter simply has a unique \irrep, but the former only has a unique \emph{regular} \irrep.
\smallskip

\noindent 
\textbf{5.} In $d=2$ the equivalence between the operator and configuration space approaches breaks down, because $S_N\neq \pi_1(Q_N)=B_N$. 
Even defining  the operator quantum theory on $H_N=L^2(\til{Q}_N)$, with algebra of observables $\mathcal{M}_N=B(L^2(\til{Q}_N))^{B_N}$, fails to rescue the equivalence, because  the decomposition of $H_N$ under $\mathcal{M}_N$ by no means contains all \irrep s of $B_N$. In this case deformation quantization gives many more sectors than the improved operator approach (which in turn gives more sectors than the naive one of \S\ref{S2}).

\end{document}